\title{NARROW ESCAPE, part III: Riemann surfaces and non-smooth domains}
\author{A. Singer \thanks{Department of Applied Mathematics,
Tel-Aviv University, Ramat-Aviv, 69978 Tel-Aviv, Israel, e-mail:
amits@post.tau.ac.il}\,,\ \ Z. Schuss\thanks{Department of
Mathematics, Tel-Aviv University, Tel-Aviv 69978,  Israel, e-mail:
schuss@post.tau.ac.il.}\,,\ \ D. Holcman\thanks{Department of
Mathematics, Weizmann Institute of Science, Rehovot 76100 Israel,
e-mail holcman@wisdom.weizmann.ac.il. }
\thanks{Keck Center, department of Physiology, UCSF, 513 Parnassus
Ave, San Francisco 94143 USA, e-mail holcman@phy.ucsf.edu.}
}
\documentclass[12pt]{article}
\usepackage{float}
\usepackage{amsmath}
\usepackage{epsfig, graphicx} 
\newcommand{\mb}[1]{\mbox{\boldmath$#1$}}
\newcommand{\p}{\partial}
\newcommand{\ds}{\displaystyle}
\newcommand{\beq}{\begin{eqnarray}}
\newcommand{\beqq}{\begin{eqnarray*}}
\newcommand{\eeq}{\end{eqnarray}}
\newcommand{\eeqq}{\end{eqnarray*}}
\newcommand{\x}{\mbox{\boldmath$x$}}
\newcommand{\y}{\mbox{\boldmath$y$}}

\newcommand{\n}{\mbox{\boldmath$n$}}

\font\bb=msbm10 at 12pt

\def\rR{\hbox{\bb R}}

\def\ds#1{\displaystyle{#1}}

\begin{document}
\numberwithin{equation}{section} \maketitle

\begin{abstract}
We consider Brownian motion in a bounded domain $\Omega$ on a
two-dimensional Riemannian manifold $(\Sigma,g)$. We assume that
the boundary $\p\Omega$ is smooth and reflects the trajectories,
except for a small absorbing arc $\p\Omega_a\subset\p\Omega$. As
$\p\Omega_a$ is shrunk to zero the expected time to absorption in
$\p\Omega_a$ becomes infinite. The narrow escape problem consists
in constructing an asymptotic expansion of the expected lifetime,
denoted $E\tau$, as $\varepsilon=|\partial \Omega_a|_g/|\partial
\Omega|_g\to0$. We derive a leading order asymptotic approximation
$E\tau = \ds{
\frac{|\Omega|_g}{D\pi}}\left[\log\ds{\frac{1}{\varepsilon}}+O(1)\right]$.
The order 1 term can be evaluated for simply connected domains on
a sphere by projecting stereographically on the complex plane and
mapping conformally on a circular disk. It can also be evaluated
for domains that can be mapped conformally onto an annulus. This
term is needed in real life applications, such as trafficking of
receptors on neuronal spines, because
$\log\ds{\frac{1}{\varepsilon}}$ is not necessarily large, even
when $\varepsilon$ is small. If the absorbing window is located at
a corner of angle $\alpha$, then $E\tau = \ds{
\frac{|\Omega|_g}{D\alpha}}\left[\log\ds{\frac{1}{\varepsilon}}+O(1)\right],$
if near a cusp, then $E\tau$ grows algebraically, rather than
logarithmically. Thus, in the domain bounded between two tangent
circles, the expected lifetime is $E\tau =
\ds{\frac{|\Omega|}{(d^{-1}-1)D}}\left(\frac{1}{\varepsilon} +
O(1) \right)$.
\end{abstract} \maketitle

\section{Introduction}
In many applications it is necessary to find the mean first
passage time (MFPT) of a Brownian particle to a small absorbing
window in the otherwise reflecting boundary of a given bounded
domain. This is the case, for example, in the permeation of ions
through protein channels of cell membranes \cite{Hille}, and in
the trafficking of AMPA receptors on nerve cell membranes
\cite{Holcman}, \cite{Choquet}. While the first example is three
dimensional the second is two dimensional, which leads to very
different results. In this paper we consider the two dimensional
case.

In the first two parts of this series of papers, we considered the
narrow escape problem in three dimensions \cite{NarrowEscape1} and
in the planar circular disk \cite{NarrowEscape2}. The leading
order asymptotic behavior of the MFPT is different in the three
and two dimensional cases; it is proportional to the relative size
of the reflecting and absorbing boundaries in three dimensions,
but in two dimensions it is proportional to the logarithm of this
quotient. The difference in the orders of magnitude is the result
of the different singularities of Neumann's function for Laplace's
equation in the two cases.

While the second term in the asymptotic expansion of the MFPT in
three dimensions is much smaller then the first one, it is not
necessarily so in two dimensions, because of the slow growth of
the logarithmic function. It is necessary, therefore, to find the
second term in the expansion in the two-dimensional case. This
term was found for the case of a planar circular disk in
\cite{NarrowEscape2}, and can therefore be found for all simply
connected domains in the plane that can be mapped conformally onto
the disk. Similarly, it can be found for simply connected domains
on two-dimensional Riemannian manifolds that can be mapped
conformally on the planar disk. For example, the sphere with a
circular cap cut off can be projected stereographically onto the
disk, and so the second term for the narrow escape problem for
such domains can be found.

The specific mathematical problem can be formulated as follows. A
Brownian particle diffuses freely in a bounded domain $\Omega$ on
a two-dimensional Riemannian manifold $(\Sigma,g)$. The boundary
$\p\Omega$ is reflecting, except for a small absorbing arc
$\p\Omega_a$. The ratio between the arclength of the absorbing
boundary and the arclength of the entire boundary is a small
parameter,
 $$\varepsilon = \ds{\frac{|\partial
 \Omega_a|_g}{|\partial \Omega|_g}} \ll 1.$$
The MFPT to $\p\Omega_a$, denoted $E\tau$, becomes infinite as
$\varepsilon\to0$.

In this paper we calculate the first term in the asymptotic
expansion of $E\tau$ for a general smooth bounded domain on a
general two-dimensional Riemannian manifold. We find the second
term for an annulus of two concentric circles, with a small hole
located on its inner boundary. This result is generalized in a
straightforward manner to domains that are conformally equivalent
to the annulus.

The calculation of the second term involves the solution of the
mixed Dirichlet-Neumann problem for harmonic functions in
$\Omega$. While in the three dimensional case this is a classical
problem in mechanics, diffusion, elasticity theory, hydrodynamics,
and electrostatics \cite{Sneddon}-\cite{Fabrikant2}, the two
dimensional problem did not draw as much attention in the
literature.

First, we consider the problem of narrow escape on two dimensional
manifolds, and derive the leading order asymptotic approximation
 \beq
E\tau = \ds{
\frac{|\Omega|_g}{D\pi}}\left[\log\ds{\frac{1}{\varepsilon}}+O(1)\right]\quad\mbox{for}\quad
\varepsilon\ll1.\label{eq:holcman}
 \eeq
This generalizes the result of \cite{Holcman} from general smooth
planar domains to general domains on general smooth
two-dimensional Riemannian manifolds.

The second term in the asymptotic expansion is found for the
2-sphere $x^2+y^2+z^2=R^2$. The calculation is made possible by
the stereographic projection that maps the Riemann sphere onto a
circular disk, a problem that was solved in \cite{NarrowEscape2}.
The boundary in this case is a spherical cap of central angle
$\delta$ at the north pole, where $\varepsilon$ is the ratio
between the absorbing arc and the entire boundary circle. We find
that the MFPT, averaged with respect to an initial uniform
distribution, is given by
\begin{equation}
E\tau = \frac{|\Omega|_g}{2\pi D}
\left[\log\frac{1}{\delta}+2\log\frac{1}{\varepsilon}+3\log 2 -
\frac{1}{2} +
O(\varepsilon,\delta^2\log\delta,\delta^2\log\varepsilon) \right],
\end{equation}
where $|\Omega|_g=4\pi R^2$ is the surface area of the sphere.
Note that there are two small parameters that control the behavior
of the MFPT in this problem. The small $\varepsilon$ contributes
as equation (\ref{eq:holcman}) predicts, whereas the small
$\delta$ parameter contributes half as much.

\begin{figure}
\includegraphics{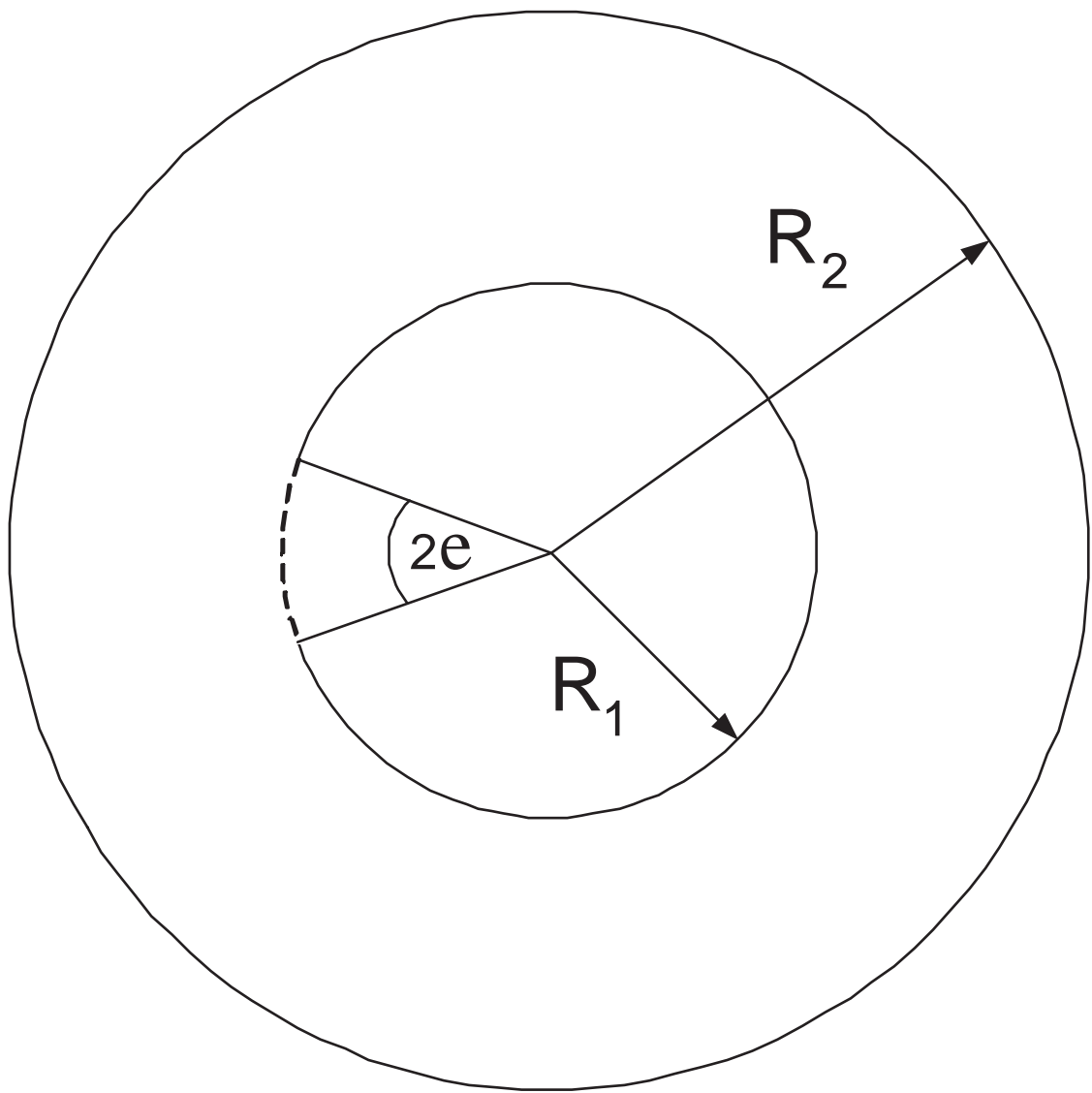}
\caption{An annulus $R_1 < r < R_2$. The particle is absorbed at
an arc of length $2\varepsilon R_1$ (dashed line) at the inner
circle. The solid lines indicate reflecting
boundaries.}\label{f:annulus}
\end{figure}

The second case that we consider is that of narrow escape from an
annulus, whose boundary is reflecting, except for a small
absorbing arc on the inner circle. Specifically, the annulus is
the domain $R_1 < r < R_2$, with all reflecting boundaries except
for a small absorbing window located at the inner circle (see Fig.
\ref{f:annulus}). The inversion $w=1/z$ transforms this case into
that of the absorbing boundary on the outer circle. Setting $\beta
= \ds{\frac{R_1}{R_2}} < 1$, the MFPT, averaged with respect to a
uniform initial distribution, can be written as
 \beqq
&&E\tau =\\
&&\nonumber\\
&& (R_2^2-R_1^2)\left[\log \frac{1}{\varepsilon} + \log 2 +
2\beta^2 \right] +
\frac{1}{2}\frac{R_2^2}{1-\beta^2}\log\frac{1}{\beta} -
\frac{1}{4}R_2^2 + O(\varepsilon,\beta^4)R_2^2.
 \eeqq
Also in this case we find two small parameters, the $\varepsilon$
contribution belongs to a singular perturbation problem with a
boundary layer solution and an almost constant outer solution with
singular fluxes near the edges of the window, whereas the $\delta$
contribution is just the singularity of Green's function at the
origin--a problem with a regular flux. This result is generalized
to a sphere with two antipodal circular caps removed. We find that
for $\beta \ll 1$ the maximum exit time is attained near the south
pole, as expected. This result can be generalized to manifolds
that can be mapped conformally onto the said domain.

\begin{figure}
\includegraphics{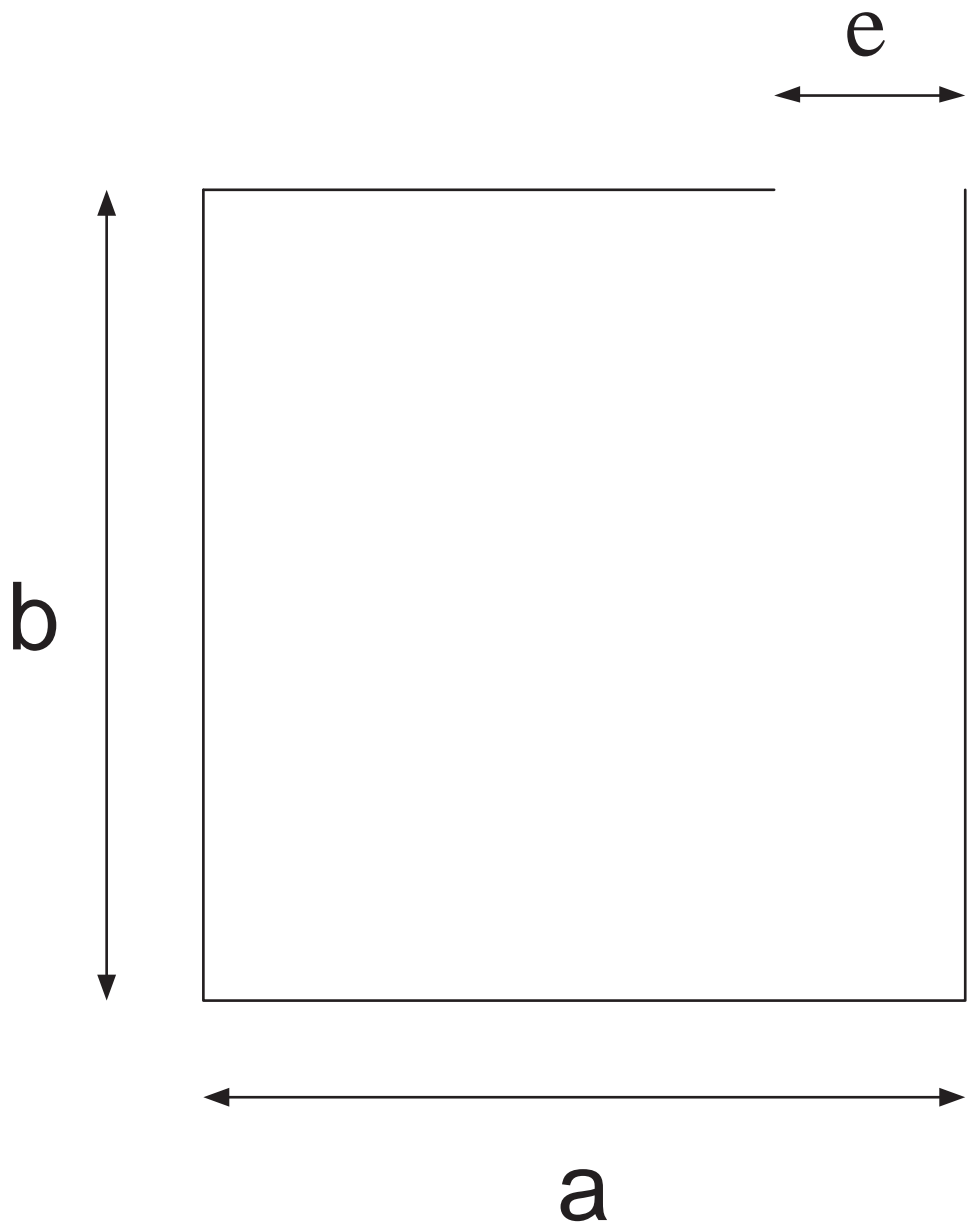}
\caption{Rectangle of sizes $a$ and $b$ with a small absorbing
segment of size $\varepsilon$ at the corner.}\label{f:rectangle}
\end{figure}

\begin{figure}
\includegraphics[width=6.5in]{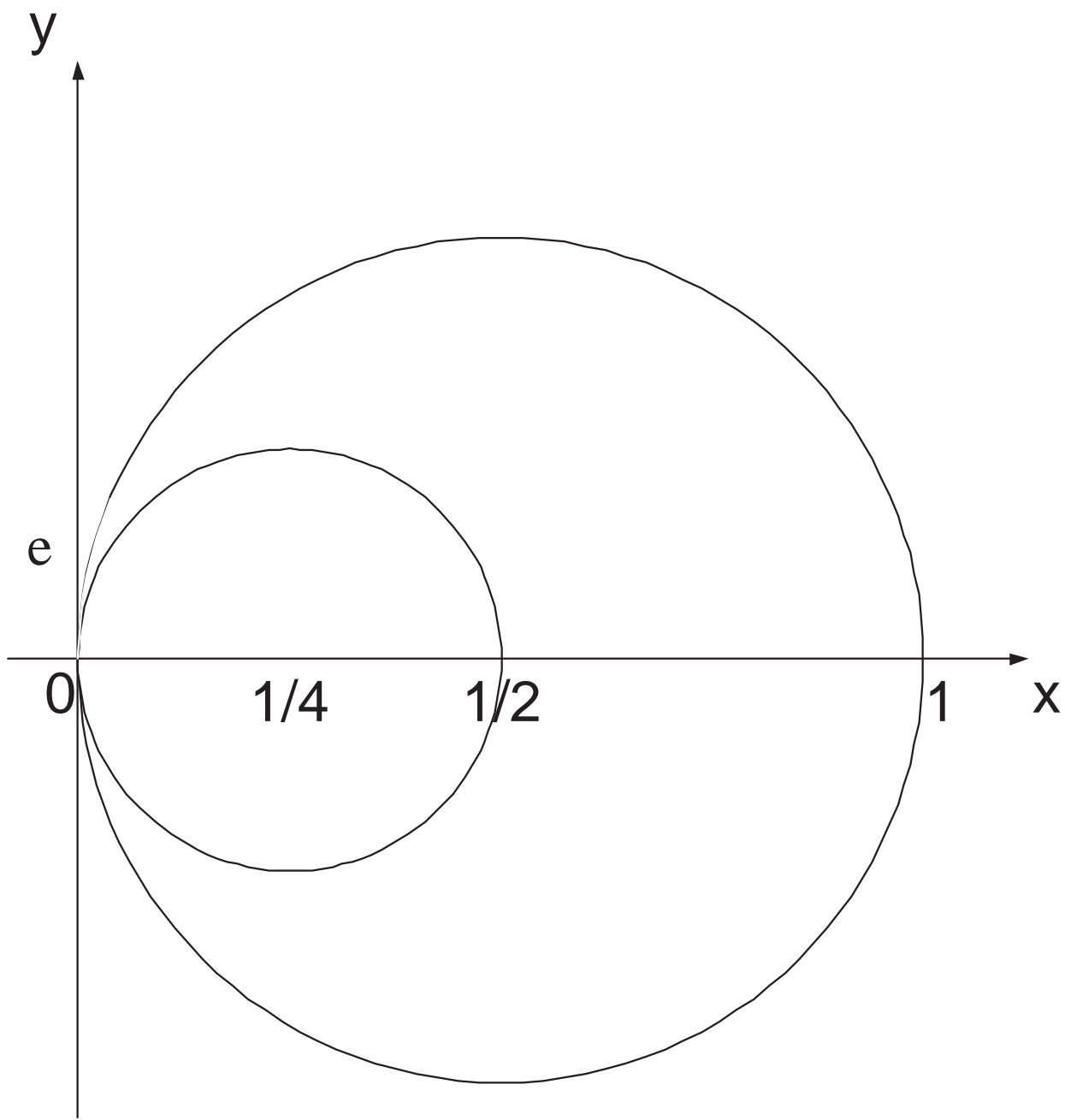}
\caption{The point $(0,0)$ is a cusp point of the dotted domain
bounded between the two circles. The small absorbing arc of length
$\varepsilon$ is located at the cusp point.}\label{f:cusp}
\end{figure}

The asymptotic expansion of the MFPT to a non-smooth part of the
boundary is different. We consider two types of singular boundary
points: corners and cusps. If the absorbing arc is located at a
corner of angle $\alpha$, the MFPT is
\begin{equation}
\label{eq:corner-MFPT} E\tau = \ds{
\frac{|\Omega|_g}{D\alpha}}\left[\log\ds{\frac{1}{\varepsilon}}+O(1)\right].
\end{equation}
For example, the MFPT from a rectangle with sides $a$ and $b$ to
an absorbing window of size $\varepsilon$ at the corner ($\alpha =
\pi / 2$, see Figure \ref{f:rectangle}), is
$$E\tau = \frac{2|\Omega|}{\pi}\left[\log \frac{a}{\varepsilon} +
\log \frac{2}{\pi} + \frac{\pi}{6}\frac{b}{a} + 2\beta^2 +
O\left(\frac{\varepsilon}{a},\beta^4\right)\right],$$ where
$|\Omega|=ab$ and $\beta=e^{-\pi b/a}$. The calculation of the
second order term turns out to be similar to that in the annulus
case. The pre-logarithmic factor $\ds{\frac{|\Omega|_g}{D\alpha}}$
is the result of the different singularity of the Neumann function
at the corner. It can be obtained by either the method of images,
or by the conformal mapping $z\mapsto z^{\pi/\alpha}$ that
flattens the corner. In the vicinity of a cusp $\alpha \to 0$,
therefore the asymptotic expansion (\ref{eq:corner-MFPT}) is
invalid. We find that near a cusp the MFPT grows algebraically
fast as $\ds{\frac{1}{\varepsilon^\lambda}}$, where $\lambda$ is
the order of the cusp. Note that the MFPT grows faster to infinity
as the boundary is more singular. The change of behavior from a
logarithmic growth to an algebraic one expresses the fact that
entering a cusp is a rare Brownian event. For example, the MFPT
from the domain bounded between two tangent circles to a small arc
at the common point (see Figure \ref{f:cusp}) is $E\tau =
\ds{\frac{|\Omega|}{(d^{-1}-1)D}}\left(\frac{1}{\varepsilon} +
O(1) \right)$, where $d<1$ is the ratio of the radii. This result
is obtained by mapping the cusped domain conformally onto the
upper half plane. The singularity of the Neumann function is
transformed as well. The leading order term of the asymptotics can
be found for any domain that can be mapped conformally to the
upper half plane.

In three dimensions the class of isolated singularities of the
boundary is much richer than in the plane. The results of
\cite{NarrowEscape1} cannot be generalized in a straightforward
way to windows located near a singular point or arc of the
boundary. We postpone the investigation of the MFPT to windows at
isolated singular points in three dimensions to a future paper.

As a possible application of the present results, we mention the
calculation of the diffusion coefficient from the statistics of
the lifetime of a receptor in a corral on the surface of a
neuronal spine \cite{Choquet}.

\section{Asymptotic approximation to the MFPT on a Riemannian manifold}
\label{sec:leading} We denote by $\x(t)$ the trajectory of a
Brownian motion in a bounded domain $\Omega$ on a two-dimensional
Riemannian manifold $(\Sigma,g)$. For a domain $\Omega \subset
\Sigma$ with a smooth boundary $\p\Omega$ (at least $C^1$), we
denote by $|\Omega|_g$ the Riemannian surface area of $\Omega$ and
by $|\p\Omega|_g$ the arclength of its boundary, computed with
respect to the metric $g$. The boundary $\p\Omega$ is partitioned
into an absorbing arc $\p\Omega_a$ and the remaining part
$\p\Omega-\p\Omega_a$ is reflecting for the Brownian trajectories.
We assume that the absorbing part is small, that is,
 $$\varepsilon =\frac{|\p\Omega_a|_g}{|\p\Omega|_g}\ll1,$$
however, $\Sigma$ and $\Omega$ are independent of $\varepsilon$;
only the partition of the boundary $\p\Omega$ into absorbing and
reflecting parts varies with $\varepsilon$.

The first passage time $\tau$ of the Brownian motion from $\Omega$
to $\p\Omega_a$ has a finite mean and we define
 \[u(\x)=E[\tau\,|\,\x(0)=\x].\]
The function $u(\x)$ satisfies the mixed Neumann-Dirichlet
boundary value problem (see for example \cite{McKean1},
\cite{Schuss})
 \beq
D\Delta_g u(\x)&=&-1\quad\mbox{for}\quad \x\in \Omega\label{D1}\\
&&\nonumber\\
\frac{\p u(\x)}{\p n}&=&0\quad\mbox{for}\quad
\x\in\p\Omega-\p\Omega_a\label{Neumann}\\
&&\nonumber\\
u(\x)&=&0\quad\mbox{for}\quad\x\in\p\Omega_a,\label{Dirichlet}
 \eeq
where $\Delta_g$ is the Laplace-Beltrami operator on $\Sigma$ and
$D$ is the diffusion coefficient. Obviously, $u(\x)\to\infty$ as
$\varepsilon\to0$, except for $\x$ in a boundary layer near
$\p\Omega_a$.

\subsection{Expression of the MFPT using the Neumann function}
We consider the Neumann function defined on $\Sigma$ by
 \beq\label{Green}
\Delta_g N(\x,\y) &=& -\delta(\x-\y)+\frac{1}{|\Omega|_g},\quad \mbox{for}\quad \x,\y \in\Omega\\
&&\nonumber\\
 \frac{ \partial N(\x,\y) }{\p \n} &=& 0,\quad
\mbox{for}\quad \x \in \p\Omega,\ \y\in\Omega. \nonumber
 \eeq
The Neumann function $N(\x,\y)$ is defined up to an additive
constant and is symmetric \cite{Garabedian}. The Neumann function
exists for the domain $\Omega$, because the compatibility
condition is satisfied (i.e., both sides of eq.(\ref{Green})
integrate to 0 over $\Omega$ due to the boundary condition). The
Neumann function $N(\x,\y)$ is constructed by using a parametrix
$H(\x,\y)$ \cite{Aubin} ,
 \beq
H(\x,\y)= -\frac{h(d(\x,\y))}{2\pi}\log d(\x,\y), \eeq where
$d(\x,\y)$ is the Riemannian distance between $\x$ and $\y$ and
$h(\cdot)$ is a regular function with compact support, equal to 1
in a neighborhood of $\y$. As a consequence of the construction
$N(\x,\y)-H(\x,\y)$ is a regular function on $\Omega$.

To derive an integral representation of the solution $u$, we
multiply eq.(\ref{D1}) by $N(\x,\y)$, eq.(\ref{Green}) by $u(\x)$,
integrate with respect to $\x$ over $\Omega$, and use Green's
formula to obtain the identity
 \beq
 \oint_{\p\Omega}N(\x(\mb{S}),\mb{\xi})\frac{\p
u(\x(\mb{S}))}{\p n}\,dS_g &=& -\frac1{|\Omega|_g}\int_{\Omega}u(\x)\,dV_g+ u(\mb{\xi})\nonumber\\
&&\nonumber\\
&& -\int_{\Omega}N(\x,\mb{\xi})\,dV_g.\label{uC}
  \eeq
The integral
 \beq
 C_\varepsilon=\ds{\frac1{|\Omega|_g}\int_{\Omega}u(\x)\,dV_g}\label{Ceps}
 \eeq
is an additive constant and the flux on the reflecting boundary
vanishes, so we rewrite eq.(\ref{uC}) as
 \beq
u(\mb{\xi})= C_\varepsilon+\int_{\Omega}N(\x,\mb{\xi})\,dV_g+
\int_{\p\Omega_a}N(\x(\mb{S}),\mb{\xi})\frac{\p u(\x(\mb{S}))}{\p
n}\,dS_g, \label{rep'}
 \eeq
where $\mb{S}$ is the coordinate of a point on $\p\Omega_a$, and
$dS_g$ is arclength element on $\p\Omega_a$ associated with the
metric $g$. Setting
 \[f(\mb{S})=\frac{\p u(\x(\mb{S}))}{\p n},\]
and choosing $\mb{\xi}\in\p\Omega_{a}$ in eq.(\ref{rep'}), we
obtain
 \beq
0= C_\varepsilon+\int_{\Omega}N(\x,\mb{\xi})\,dV_g+
\int_{\p\Omega_a}N(\x(\mb{S}),\mb{\xi})f(\mb{S})\,dS_g.
\label{Seq'}
 \eeq
The first integral in eq.(\ref{rep'}) is a constant (independent
of $\varepsilon$), because due to the  symmetry of $N(\x,\y)$
eq.(\ref{Green}) gives the boundary value problem
 \beqq
 \Delta_{\mb{\xi}}\int_{\Omega}N(\x,\mb{\xi})\,dV_g&=&0\quad\mbox{for}\quad
\mb{\xi}\in \Omega\label{intN}\\
&&\nonumber\\
 \frac{\p}{\p
n(\mb{\xi})}\int_{\Omega}N(\x,\mb{\xi})\,dV_g&=&0\quad\mbox{for}\quad
\mb{\xi}\in\p\Omega,\label{BCintN}
 \eeqq
whose solution is any constant. Changing the definition of the
constant $C_\varepsilon$,  equation (\ref{rep'}) can be written
as,
 \beq
u(\mb{\xi})= \int_{\p \Omega_a}N(\x(\mb{S}),\mb{\xi})
f(\mb{S})\,dS_g+C_\varepsilon,\label{rep}
 \eeq
and both $f(\mb{S})$ and $C_\varepsilon$ are determined by the
absorbing condition (\ref{Dirichlet})
 \beq
0=\int_{\p \Omega_a}N(\x(\mb{S}),\mb{\xi})
f(\mb{S})\,dS_g+C_\varepsilon
\quad\mbox{for}\quad\mb{\xi}\in\p\Omega_a .\label{Seq}
 \eeq
Equation (\ref{Seq}) has been considered in \cite{Holcman} for a
domain $\Sigma\subset \rR^2$ as an integral equation for
$f(\mb{S})$ and $C_\varepsilon$.

Actually, the boundary coordinate $\mb{S}$ can be chosen as
arclength on $\p\Omega_a$, denoted $s$. Under the regularity
assumptions of the boundary, the normal derivative $f(s)$ is a
regular function, but develops a singularity as $\mb{\xi}(s)$
approaches the corner boundary of $\p\Omega_a$ in $\p\Omega$
\cite{Mazya1}. Both can be determined from the representation
(\ref{rep}), if all functions in eq.(\ref{Seq}) and the boundary
are analytic. In that case the solution has a series expansion in
powers of arclength on $\Omega_a$. The method to compute
$C_\varepsilon$ follows the same step as in \cite{Holcman}.

\subsection{Leading order asymptotics}\label{small}
Under our assumptions, $u(\mb{\xi})\to\infty$ as $\varepsilon\to0$
for any fixed $\mb{\xi}\in\Omega$, so that eq.(\ref{Ceps}) implies
that $C_\varepsilon\to\infty$ as well. It follows from
eq.(\ref{Seq}) that the integral in (\ref{Seq}) decreases to
$-\infty$.

An origin $0 \in \p\Omega_a$ is fixed and the boundary $\p\Omega$
is parameterized by $(x(s),y(s))$.  We rescale $s$ so that
$\p\Omega=\left\{(x(s),y(s))\,:\,-\frac12<s\leq\frac12\right\}$
and $\left(x\left(-\frac12\right),y\left(-\frac12\right)\right)=
\left(x\left(\frac12\right),y\left(\frac12\right)\right)$. We
assume that the functions $x(s)$ and $y(s)$ are real analytic in
the interval $2|s|<1$ and that the absorbing part of the boundary
$\p \Omega_a$ is the arc
\[\p\Omega_a=\left\{(x(s),y(s))\,:\,|s|<\varepsilon\right\}.\]
The Neumann function can be written as
 \beq
N(\x,\mb{\xi})=-\frac{1}{2\pi}\log
d(\x,\mb{\xi})+v_N(\x,\mb{\xi}),\label{Nf}\quad \hbox{for}\quad \x
\in B_{\delta}(\mb{\xi}),
 \eeq
where $ B_{\delta}(\mb{\xi})$ is a geodesic ball of radius
$\delta$ centered at $\mb{\xi}$ and $v_N(\x;\mb{\xi})$ is a
regular function. We consider a normal geodesic coordinate system
$(x,y)$ at the origin, such that one of the coordinates coincides
with the tangent coordinate to $\p\Omega_a$. We choose unit
vectors $\mb{e}_1,\mb{e}_2$ as an orthogonal basis in the tangent
plane at 0 so that for any vector field
$\mb{X}=x_1\mb{e}_1+x_2\mb{e}_2$, the metric tensor $g$ can be
written as
 \beq
g_{ij}= \delta_{ij}+\varepsilon^2 \sum_{kl}a^{kl}_{ij}x_k x_l
+o(\varepsilon^2),
 \eeq
where $|x_k|\leq 1$, because $\varepsilon$ is small. It follows
that for $\x,\y$ inside the geodesic ball or radius $\varepsilon$,
centered at the origin, $d(\x,\y)=d_E(\x,\y) +O(\varepsilon^2)$,
where $d_E$ is the Euclidean metric. We can now use the
computation given in the Euclidean case in \cite{Holcman}. To
estimate the solution of equation (\ref{Seq}), we recall that when
both $\x$ and $\mb{\xi}$ are on the boundary, $v_N(\x,\mb{\xi})$
becomes singular (see  \cite[p.247, eq.(7.46)]{Garabedian}) and
the singular part gains a factor of 2, due to the singularity of
the ``image charge". Denoting by $ \tilde v_N$ the new regular
part, equation (\ref{Seq}) becomes
 \beq \label{eqqexpan} \int_{|s'|<\varepsilon} \left[
\tilde v_N(\x(s');\mb{\xi}(s))- \frac{\log d(\x(s),\mb{\xi}(s'))
}{\pi}\right]f(s')\,S(ds') =C_\varepsilon,\label{Ceq}
 \eeq
where $S(ds')$ is the induced measure element on the boundary, and
$\x=(x(s),y(s))$, $\mb{\xi} =(\xi(s),\eta(s))$. Now, we expand the
integral in eq.(\ref{eqqexpan}), as in \cite{Holcman},
\[
 \log d(\x(s),\mb{\xi}(s')) =  \log\left(\sqrt{(x(s')-\xi(s))^2+ (y(s')-\eta(s))^2}\,
 \left(1 +O(\varepsilon^2)\right)\right)
\]
and
 \beq
S(ds) f(s)=\sum_{j=0}^\infty f_j s^j\,ds,\quad \tilde
v_N(\x(s');\mb{\xi}(s))S(ds')=\sum_{j=0}^\infty
v_j(s')s^j\,ds'\label{gexp}
 \eeq
for $|s|<\varepsilon$, where $v_j(s')$ are known coefficients and
$f_j$ are unknown coefficients, to be determined from
eq.(\ref{Ceq}). To expand the logarithmic term in the last
integral in eq.(\ref{Ceq}), we recall that $x(s'),y(s'),\xi(s)$,
and $\eta(s)$ are analytic functions of their arguments in the
intervals $|s|<\varepsilon$ and $|s'|<\varepsilon$, respectively.
Therefore
 \beq
&&\int_{-\varepsilon}^{\varepsilon}(s')^n \log
d(\x(s),\mb{\xi}(s')) \,ds'=\label{exp1}\\
&&\nonumber\\
&&\int_{-\varepsilon}^{\varepsilon}(s')^n\log\sqrt{(x(s')-\xi(s))^2+
 (y(s')-\eta(s))^2}\left(1 +O(\varepsilon^2)\right)\,ds'=\nonumber\\
 &&\nonumber\\
 &&\int_{-\varepsilon}^{\varepsilon}(s')^n
 \log\left\{|s'-s|\left(1+O\left((s'-s)^2\right)\right)\right\}\left(1 +O(\varepsilon^2)\right)\,ds'.\nonumber
 \eeq
We keep in Taylor's expansion of
$\log\left\{|s'-s|\left(1+O\left((s'-s)^2\right)\right)\right\}$
only the leading term, because higher order terms contribute
positive powers of $\varepsilon$ to the series
\begin{eqnarray}
\int_{-\varepsilon }^{\varepsilon }\log (s -s')^{2}\,ds'
=4\varepsilon \left( \log\varepsilon-1\right) +2\sum_{j=1}^{\infty }\frac{%
1}{(2j-1)j}\frac{s^{2j}}{\varepsilon ^{2j-1}}.\label{exp2}
\end{eqnarray}
For even $n\geq0$, we have
\begin{eqnarray}
&&\int_{-\varepsilon }^{\varepsilon }(s')^n\log (s -s')^{2}\,ds'
=\nonumber\\
&&\nonumber\\
&&4\left( \frac{\varepsilon ^{n+1}}{n+1}\log \varepsilon
-\frac{\varepsilon
^{n+1}}{(n+1)^{2}}\right) -2\sum_{j=1}^{\infty }s^{2j}\frac{%
\varepsilon ^{n-2j+1}}{j(n-2j+1)},\label{exp3}
\end{eqnarray}
whereas for odd $n$, we have
\begin{eqnarray}
\int_{-\varepsilon }^{\varepsilon }(s')^{n}\log (s-s')^{2}\,ds'
=-4\sum_{j=1}^{\infty }\frac{s^{2j+1}}{2j+1}\frac{\varepsilon
^{n-2j}}{n-2j}.\label{exp4}
\end{eqnarray}
Using the above expansion, we rewrite  eq.(\ref{Ceq}) as
 \beqq
0&=& \int_{-\varepsilon}^{\varepsilon}\left\{\frac{-1}{\pi}
\log\left[|s'-s|^2\left(1+O\left((s'-s)^2\right)\right)(1
+O(\varepsilon^2))\right]+ \sum_{j=0}^\infty
v_j(s')s^j\right\}\times\\
&&\\
&& \sum_{j=0}^\infty f_js'^j\, ds'+C_\varepsilon,
 \eeqq
and expand in powers of $s$. At the leading order, we obtain
 \beq
 &&\varepsilon \left( \log \varepsilon-1\right)f_0  +\sum_p
\left( \frac{\varepsilon ^{2p+1}}{2p+1}\log \varepsilon
-\frac{\varepsilon ^{2p+1}}{(2p+1)^{2}}\right)f_{2p}
=\nonumber\\
&&\nonumber\\
&&\frac{\pi}{2}\int_{-\varepsilon}^{\varepsilon} v_0(s')\,ds'
+C_\varepsilon. \label{deg0}
 \eeq
Equation (\ref{deg0}) and
 \beqq
 \frac{1}{2}\int_{-\varepsilon}^{\varepsilon} f(s)\,S(ds) =
\sum_p\frac{\varepsilon^{2p+1}}{(2p+1)}f_{2p}
 \eeqq
determine the leading order term in the expansion of
$C_\varepsilon$. Indeed, integrating eq.(\ref{D1}) over the domain
$\Omega$, we see that the compatibility condition gives
 \beq
 \int_{-\varepsilon}^\varepsilon
 f(s)\,S(ds)=-|\Omega|_g,
 \eeq
and using the fact that $\ds\int_{-\varepsilon}^{\varepsilon}
v_0(s')S(ds')=O(\varepsilon)$, we find that the leading order
expansion of $C_\varepsilon$ in eq.(\ref{deg0}) is
 \beq
 C_\varepsilon=\frac{|\Omega|_g}{\pi}\left[\log\frac1\varepsilon+O(1)\right]\quad
\mbox{for $\varepsilon\ll1$.}\label{C}
 \eeq
If the diffusion coefficient is $D$, eq.(\ref{rep}) gives the MFPT
from a point $\x \in\Omega$, outside the boundary layer, as
 \beq
E[\tau\,|\,\x]= u(\x)= \frac{|\Omega|_g}{ \pi
D}\left[\log\frac1\varepsilon+O(1)\right]\quad \mbox{for
$\varepsilon\ll1$.}\label{MFPTC}
 \eeq

\section{The annulus problem}
\label{sec:annulus} We consider a Brownian particle that is
confined in the annulus $R_1 < r < R_2$. The particle can exit the
annulus through a narrow opening of the inner circle (see
Fig.\ref{f:annulus}). The MFPT $v(\mb{x})$ satisfies
\begin{eqnarray}
\Delta v &=& -1, \quad\mbox{for}\quad R_1 < r < R_2,\label{-1} \\
&&\nonumber\\
\frac{\partial v}{\partial r} &=& 0, \quad\mbox{for}\quad r=R_2, \nonumber \\
&&\nonumber\\
 \frac{\partial v}{\partial r} &=& 0, \quad\mbox{for}\quad r=R_1,
\;|\theta-\pi|>\varepsilon, \nonumber \\
&&\nonumber\\
 v&=&0, \quad\mbox{for}\quad r=R_1, \;|\theta-\pi| < \varepsilon. \nonumber
\end{eqnarray}
The function $w=\ds{\frac{R_1^2-r^2}{4}}$ is a solution of the
Dirichlet problem for eq.(\ref{-1}) in the exterior domain of the
inner circle $r>R_1$. More specifically, it satisfies the boundary
value problem
\begin{eqnarray}
\Delta w &=& -1, \quad\mbox{for}\quad R_1<r<R_2, \nonumber \\
&&\nonumber\\
 \frac{\partial w}{\partial r} &=& -\frac{1}{2}R_1,
\quad\mbox{for}\quad r=R_1,\nonumber \\
&&\nonumber\\
 \frac{\partial w}{\partial r} &=& -\frac{1}{2}R_2,
\quad\mbox{for}\quad r=R_2,\nonumber \\
&&\nonumber\\
 w&=& 0, \quad\mbox{for}\quad r=R_1.
\end{eqnarray}
The function $u=v-w$ satisfies
\begin{eqnarray}
\Delta u &=& 0, \quad\mbox{for}\quad R_1 < r < R_2, \nonumber \\
&&\nonumber\\
\frac{\partial u}{\partial r} &=& \frac{1}{2}R_2,
\quad\mbox{for}\quad r=R_2,\nonumber \\
&&\nonumber\\
\frac{\partial u}{\partial r} &=& \frac{1}{2}R_1,
\quad\mbox{for}\quad r=R_1, \; |\theta-\pi|
> \varepsilon, \nonumber \\
&&\nonumber\\
u &=& 0, \quad\mbox{for}\quad r=R_1, \; |\theta-\pi| <
\varepsilon.
\end{eqnarray}
Separation of variables produces the solution
\begin{equation}
u(r,\theta) = \frac{a_0}{2} + \sum_{n=1}^\infty \left[a_n
\left(\frac{r}{R_2}\right)^n + b_n \left(\frac{R_2}{r}\right)^{n}
\right]\cos n\theta + \alpha \log\left(\frac{r}{R_1} \right),
\end{equation}
where $a_n,\,b_n$ and $\alpha$ are to be determined by the boundary
conditions. Differentiating with respect to $r$ yields
\begin{equation}
\frac{\partial u}{\partial r} = \sum_{n=1}^\infty
n\left[\frac{a_n}{R_2}\left(\frac{r}{R_2}\right)^{n-1} - \frac{b_n
R_2}{r^2} \left(\frac{R_2}{r} \right)^{n-1}\right]\cos n\theta +
\frac{\alpha}{r}.
\end{equation}
Setting $r=R_2$ gives
\begin{equation}
\frac{1}{2}R_2 = \frac{1}{R_2}\left[\sum_{n=1}^\infty n\left(a_n -
b_n \right)\cos n\theta + \alpha\right],
\end{equation}
therefore, $a_n=b_n$ and $\alpha=\ds{\frac{1}{2}}R_2^2$, and we
have
\begin{equation}
u(r,\theta) = \frac{a_0}{2} + \sum_{n=1}^\infty a_n \left[
\left(\frac{r}{R_2}\right)^n + \left(\frac{R_2}{r}\right)^{n}
\right]\cos n\theta + \frac{1}{2}R_2^2 \log\left(\frac{r}{R_1}
\right).
\end{equation}
The boundary conditions at $r=R_1$ become the dual series
equations
\begin{eqnarray}
\frac{a_0}{2} + \sum_{n=1}^\infty a_n \left[\left(\frac{R_2}{R_1}
\right)^n + \left(\frac{R_1}{R_2} \right)^n \right]\cos n \theta
&=& 0, \quad\mbox{for}\quad |\theta-\pi| < \varepsilon, \nonumber \\
&&\nonumber\\
\sum_{n=1}^\infty na_n \left[\left(\frac{R_2}{R_1} \right)^{n+1} -
\left(\frac{R_1}{R_2} \right)^{n-1}\right]\cos n\theta &=&
\frac{R_2}{2R_1}\left(R_2^2-R_1^2 \right), \quad\mbox{for}\quad
|\theta -\pi| > \varepsilon.\nonumber
\end{eqnarray}
Setting
\begin{equation}
c_n = \frac{R_1}{R_2} \left[\left(\frac{R_2}{R_1}\right)^{n+1} -
\left(\frac{R_1}{R_2} \right)^{n-1} \right]a_n, \quad \mbox{for
}\quad n\geq1,
\end{equation}
and $c_0 = a_0$ converts the dual series equations to
\begin{eqnarray}
\frac{c_0}{2} + \sum_{n=1}^\infty \frac{c_n}{1+H_n} \cos n\theta
&=& 0,\quad\mbox{for}\quad \pi-\varepsilon < \theta < \pi, \\
&&\nonumber\\
 \sum_{n=1}^\infty nc_n \cos n\theta &=&
\frac{1}{2}(R_2^2-R_1^2), \quad\mbox{for}\quad 0 < \theta < \pi-
\varepsilon, \label{eq:c_n-ref}
\end{eqnarray}
where $H_n=\ds -\frac{2\beta^{2n}}{1+\beta^{2n}}$ for $n\geq 1$,
and $H_0=0$, with $\beta=\ds{\frac{R_1}{R_2}} < 1$. Note that $H_n
= O(\beta^{2n})$ which tends to zero exponentially fast (much
faster than the $n^{-1}$ decay required for the Collins method
\cite{Collins1,Collins2}, see also \cite{NarrowEscape1}).

The case $H_n \equiv 0$ was solved in \cite{NarrowEscape2}. We now
try to find the correction of that result due to the non vanishing
$H_n$. As in \cite{NarrowEscape2} the equation
 \beqq
&&\frac{c_0}{2} + \sum_{n=1}^\infty \frac{c_n}{1+H_n} \cos
n\theta=\\
&&\\
&&\cos \frac{\theta}{2} \int_{\theta}^{\pi-\varepsilon}
\frac{h_1(t)\,dt}{\sqrt{\cos \theta - \cos
t}}\quad\mbox{for}\quad\mbox{for}\quad 0<\theta<\pi-\varepsilon
 \eeqq
defines the function $h_1(\theta)$ uniquely for
$0<\theta<\pi-\varepsilon$, the coefficients are given by
\begin{equation}\label{eq:c-h}
c_n = \frac{1+H_n}{\sqrt{2}} \int_0^{\pi-\varepsilon} h_1(t)
\left[P_n(\cos t) + P_{n-1}(\cos t) \right]\,dt,
\end{equation}
and
\begin{equation}
\label{eq:c_0-1} c_0 = \sqrt{2} \int_0^{\pi-\varepsilon}
h_1(t)\,dt.
\end{equation}
Integrating equation (\ref{eq:c_n-ref}) gives
\begin{equation}
\sum_{n=1}^\infty c_n \sin n \theta = \frac{1}{2}\left(R_2^2-R_1^2
\right)\theta, \quad \mbox{for}\quad 0 < \theta < \pi -
\varepsilon.\label{sumn}
\end{equation}
Substituting eq.(\ref{sumn}) in equation (\ref{eq:c-h}), changing
the order of summation and integration, while using
\cite[eq.(2.6.31)]{Sneddon},
\begin{equation}
\label{eq:heavyside} \frac{1}{\sqrt{2}}\sum_{n=1}^\infty [P_n(\cos
t) + P_{n-1}(\cos t)]\sin n\theta = \frac{\cos\frac{1}{2}\theta
H(\theta-t)}{\sqrt{\cos t - \cos \theta}},
\end{equation}
we obtain for $0 < \theta < \pi - \varepsilon,$
 \beq
\int_0^{\theta} \frac{h_1(t)}{\sqrt{\cos t - \cos \theta}}\,dt +
\int_0^{\pi-\varepsilon} K_{\beta}(\theta,t)h_1(t)\,dt=
\frac{\left(R_2^2-R_1^2 \right)\theta}{2\cos\ds{
\frac{\theta}{2}}},\label{eq:abel+K}
 \eeq
where the kernel $K_\beta$ is
\begin{eqnarray}
K_{\beta}(\theta,t) &=& \frac{1}{\sqrt{2}\cos
\ds{\frac{\theta}{2}}} \sum_{n=1}^\infty H_n \left(P_n(\cos t) +
P_{n-1}(\cos t) \right)
\sin n \theta \nonumber \\
&&\nonumber\\
 &=& -2\sqrt{2}(1+\cos t)\sin\frac{\theta}{2}
\,\beta^2 + O(\beta^4).\label{Kb}
\end{eqnarray}
The infinite sum in eq.(\ref{Kb}) is approximated by its first
term, while using the first two Legendre polynomials $P_0(x)=1,\;
P_1(x)=x$. Using Abel's inversion formula applied to equation
(\ref{eq:abel+K}), we find that
\begin{equation}
\label{eq:fredholm-second-type} h_1(t) - \int_0^{\pi-\varepsilon}
\tilde{K}_\beta(t,s)h_1(s)\,ds =
\frac{R_2^2-R_1^2}{\pi}\frac{d}{dt} \int_0^t \frac{u\sin
\ds{\frac{u}{2}}}{\sqrt{\cos u - \cos t}}\,du,
\end{equation}
where the kernel $\tilde{K}_\beta$ is
\begin{eqnarray}
\tilde{K}_\beta(t,s) &=& -\frac{1}{\pi} \frac{d}{dt} \int_0^t
\frac{K_\beta (u,s)\sin u}{\sqrt{\cos u - \cos t}} \,du
\\
&&\nonumber\\
&=& \beta^2\,\frac{2\sqrt{2}(1+\cos s)}{\pi} \frac{d}{dt} \int_0^t
\frac{\sin\ds{\frac{u}{2}} \,\sin u}{\sqrt{\cos u - \cos t}}
\,du+ O(\beta^4).\nonumber
\end{eqnarray}
The substitution
\begin{equation}
\label{eq:magic-substitution2} s=\sqrt{\frac{\cos u - \cos t}{2}}
\end{equation}
gives
\begin{equation}
\int_0^t \frac{\sin\ds{\frac{u}{2}} \,\sin u }{\sqrt{\cos u - \cos
t}} \,du=\frac{\pi}{\sqrt{2}} \sin^2\frac{t}{2},
\end{equation}
therefore,
\begin{equation}
\tilde{K}_{\beta}(t,s) = 2\beta^2 \cos^2\frac{s}{2} \sin t +
O(\beta^4).
\end{equation}

Equation (\ref{eq:fredholm-second-type}) is a Fredholm integral
equation of the second kind for $h_1$, of the form
\begin{equation}
(I-\tilde{K}_\beta)h = z,
\end{equation}
where
 $$z(t) = \ds{\frac{R_2^2-R_1^2}{\pi}\frac{d}{dt} \int_0^t
 \frac{u\sin \ds{\frac{u}{2}}}{\sqrt{\cos u - \cos t}}}\,du.$$
Therefore, we $h$ can be expanded as
\begin{equation}
h = z + \tilde{K}_\beta z + \tilde{K}_\beta^2 z + \ldots,
\end{equation}
which converges in $L^2$. Since $c_0 = \sqrt{2} \langle h,1
\rangle$ (eq.(\ref{eq:c_0-1})), we find an asymptotic expansion of
the form
\begin{equation}
c_0 = \sqrt{2}\left[\langle z, 1 \rangle + \langle \tilde{K}_\beta
z, 1 \rangle + \ldots \right].
\end{equation}
The leading order term of this expansion was calculated in
\cite{NarrowEscape2}. We now estimate the error term $\langle
\tilde{K}_\beta z, 1 \rangle$, which is also the $O(\beta^2)$
correction. Integrating by parts and changing the order of
integration yields
\begin{eqnarray}
\tilde{K}_{\beta} z(t) &=& 2\beta^2\frac{R_2^2-R_1^2}{\pi}\sin t
\int_0^\pi \cos^2\frac{s}2\,ds\,\frac{d}{ds} \int_0^s
\frac{u\sin\ds{\frac{u}{2}}}{\sqrt{\cos u - \cos s}}\,du \nonumber \\
&&\nonumber\\
 &=& \beta^2\frac{R_2^2-R_1^2}{\pi}\sin t \int_0^\pi
\sin s \,ds \int_0^s \frac{u\sin \ds{\frac{u}{2}}\,du}{\sqrt{\cos
u -\cos s}}\nonumber \\
&&\nonumber\\ &=& \sqrt{2}\beta^2(R_2^2-R_1^2)\sin t.
\end{eqnarray}
Therefore,
\begin{equation}
\langle \tilde{K}_\beta z, 1 \rangle =
\sqrt{2}\beta^2(R_2^2-R_1^2)\int_0^\pi \sin t \,dt =
2\sqrt{2}\beta^2(R_2^2-R_1^2).
\end{equation}
We conclude that
\begin{equation}
\label{eq:c_0-annulus} c_0 = (R_2^2-R_1^2)\left[2\log
\frac{1}{\varepsilon} + 2\log 2 + 4\beta^2 +
O(\varepsilon,\beta^4)\right].
\end{equation}
The MFPT averaged with respect to a uniform initial distribution
is
\begin{eqnarray}
E\tau &=& \frac{c_0}{2} + \frac{1}{2}\frac{R_2^4}{R_2^2-R_1^2}\log
\frac{R_2}{R_1} - \frac{1}{4}R_2^2 \label{Etaub}\\
&&\nonumber\\
 &=& (R_2^2-R_1^2)\left[\log \frac{1}{\varepsilon} +
\log 2 + 2\beta^2 \right] +
\frac{1}{2}\frac{R_2^2}{1-\beta^2}\log\frac{1}{\beta} -
\frac{1}{4}R_2^2 + O(\varepsilon,\beta^4)R_2^2.\nonumber
\end{eqnarray}
Note that there are two different logarithmic contributions to the
MFPT. The ``narrow escape" small parameter $\varepsilon$
contributes
\begin{equation}
\label{eq:asym-eps} \frac{|\Omega|_g}{\pi}
\log\frac{1}{\varepsilon},
\end{equation}
as expected from the general theory (equation (\ref{eq:holcman})),
whereas the parameter $\beta$ contributes
\begin{equation}
\label{eq:asym-delta} \frac{|\Omega|_g}{2\pi} \log\frac{1}{\beta}.
\end{equation}
These asymptotics differ by a factor 2, because they account for
different singular behaviors. The asymptotic expansion
(\ref{eq:asym-eps}) comes out from a singular perturbation problem
with singular flux near the edges, boundary layer and an outer
solution, whereas the asymptotics (\ref{eq:asym-delta}) is an
immediate result of the singularity of the Neumann function, with
a regular flux.

The exit problem in an annulus with the absorbing window located
at the outer circle is solved by applying the complex inversion
mapping $z\to 1/z$, which maps the annulus into itself and
replaces the roles of the inner and outer circles. In such case,
in the limit $\beta\to 0$, the circular disk problem is recovered,
where it was shown that the maximum exit time is attained at the
antipode point on the outer circle. Since the inversion mapping
exchanges the inner and outer circles, we conclude that for the
original annulus problem, where the absorbing boundary is located
at the inner circle, the maximal exit time is attained at the
inner circle, at the antipode of the center of the hole. This
point is close by to the hole itself, a result which is somewhat
counterintuitive. Note that (\ref{Etaub}) is valid, with the
obvious modifications, for any domain that is conformally
equivalent to the annulus.

\section{Domains with corners}
\begin{figure}
\includegraphics{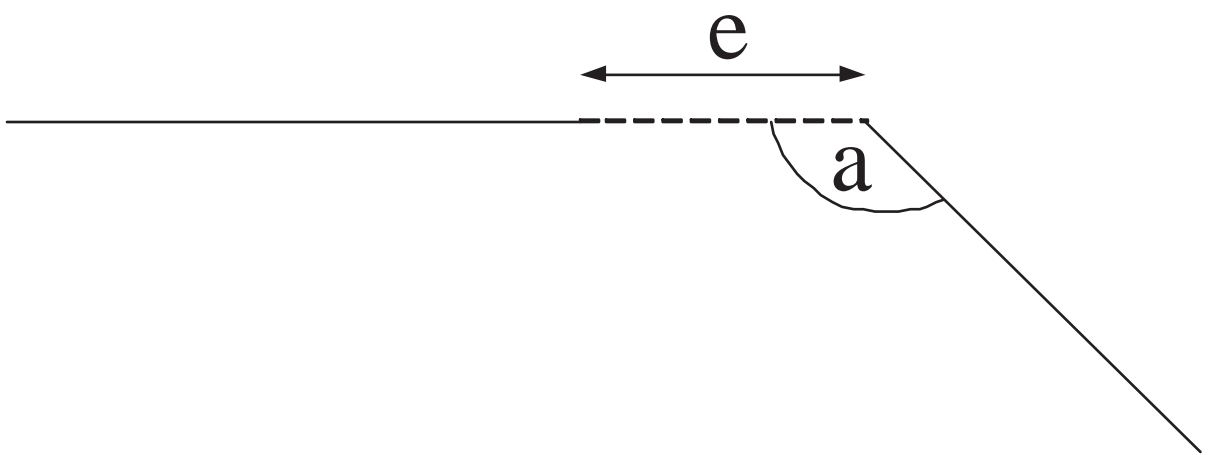}
\caption{A small opening near a corner of angle
$\alpha$.}\label{f:angle}
\end{figure}

Consider a Brownian motion in a rectangle $\Omega=(0,a)\times (0,b)$
of area $ab$. The boundary is reflecting except the small absorbing
segment $\partial \Omega_a =[a-\varepsilon,a]\times\{b \}$ (see Fig.
\ref{f:rectangle}). The MFPT $v(x,y)$ satisfies the boundary value
problem
\begin{eqnarray}
\Delta v &=& -1, \quad (x,y)\in \Omega, \nonumber \\
v &=& 0, \quad (x,y)\in \partial \Omega_a, \nonumber \\
\frac{\partial v}{\partial n} &=& 0, \quad (x,y)\in \partial \Omega
- \partial \Omega_a.
\end{eqnarray}
The function $f = \ds{\frac{b^2-y^2}{2}}$ satisfies
\begin{eqnarray}
\Delta f &=& -1, \quad (x,y)\in \Omega, \nonumber \\
f &=& 0, \quad (x,y) \in \partial \Omega_a, \nonumber \\
\frac{\partial f}{\partial n} &=& 0, \quad
(x,y)\in\{0\}\times[0,b]\cup \{a\}\times[0,b] \cup [0,a]\times\{0
\}, \nonumber \\
\frac{\partial f}{\partial n} &=& -b, \quad (x,y)\in
[0,a-\varepsilon]\times\{b \},
\end{eqnarray}
therefore, the function $u=v-f$ satisfies
\begin{eqnarray}
\Delta u &=& 0, \quad (x,y)\in \Omega, \nonumber \\
u &=& 0, \quad (x,y) \in \partial \Omega_a, \nonumber \\
\frac{\partial u}{\partial n} &=& 0, \quad
(x,y)\in\{0\}\times[0,b]\cup \{a\}\times[0,b] \cup [0,a]\times\{0
\}, \nonumber \\
\frac{\partial u}{\partial n} &=& b, \quad (x,y)\in
[0,a-\varepsilon]\times\{b \}.
\end{eqnarray}
A solution for $u$ in the form of separation of variables is
\begin{equation}
u(x,y) = \frac{a_0}{2} + \sum_{n=1}^\infty a_n \cosh \frac{\pi
ny}{a} \cos \frac{\pi n x}{a},
\end{equation}
where the coefficients $a_n$ are to be determined by the boundary
conditions at $y=b$
\begin{eqnarray}
u(x,b) &=& \frac{a_0}{2} + \sum_{n=1}^\infty a_n \cosh \frac{\pi
nb}{a} \cos \frac{\pi n x}{a} = 0, \quad x\in (a-\varepsilon,a),
\nonumber \\
\frac{\partial u}{\partial y}(x,b) &=& \frac{\pi
}{a}\sum_{n=1}^\infty na_n \sinh \frac{\pi nb}{a} \cos \frac{\pi n
x}{a} = b, \quad x\in (0,a-\varepsilon).
\end{eqnarray}
Setting $c_n = a_n \sinh \ds{\frac{\pi n b}{a}}$, we have
\begin{eqnarray}
\frac{c_0}{2} + \sum_{n=1}^\infty \frac{c_n}{1+H_n} \cos n\theta
&=& 0, \quad \pi-\delta<\theta<\pi, \nonumber \\
\sum_{n=1}^\infty nc_n \cos n\theta &=& \frac{ab}{\pi},\quad
0<\theta < \pi-\delta,
\end{eqnarray}
where $\delta = \ds{\frac{\pi \varepsilon}{a}}$ and $H_n =
\tanh\left(\ds{\frac{\pi n b}{a}} \right)-1, n\geq 1$. Note that
$H_n = O\left(\beta^{2n}\right)$ for $\beta =
\exp\left\{-\ds{\frac{\pi b}{a}}\right\} < 1$. The rectangle problem
and annulus problem (eq. (\ref{eq:c_n-ref})) are almost
mathematically equivalent, and equation (\ref{eq:c_0-annulus}) gives
the value of $c_0$
\begin{eqnarray}
c_0 &=& \frac{2ab}{\pi}\left[2\log \frac{1}{\delta} + 2\log 2 +
4\beta^2 + O(\delta,\beta^4)\right] \nonumber \\
&=& \frac{4ab}{\pi}\left[\log \frac{a}{\varepsilon} + \log
\frac{2}{\pi} + 2\beta^2 +
O\left(\frac{\varepsilon}{a},\beta^4\right)\right].
\end{eqnarray}
The error term due to $O(\beta^4)$ is generally small. For example,
in a square $a=b$ and $\beta=e^{-\pi}$ so that $\beta^4\approx
3\times 10^{-6}$. The MFPT averaged with respect to a uniform
initial distribution is
\begin{equation}
E\tau = \frac{c_0}{2} + \frac{b^2}{3} = \frac{2ab}{\pi}\left[\log
\frac{a}{\varepsilon} + \log \frac{2}{\pi} +
\frac{\pi}{6}\frac{b}{a} + 2\beta^2 +
O\left(\frac{\varepsilon}{a},\beta^4\right)\right].
\end{equation}
The leading order term of the MFPT is
\begin{equation}
\frac{2|\Omega|}{\pi}\log\frac{a}{\varepsilon},
\end{equation}
which is twice as large than (\ref{eq:asym-eps}). The general
result (\ref{eq:holcman}) was proved for a domain with smooth
boundary (at least $C^1$). However, in the rectangle example, the
small hole is located at the corner. The additional factor 2 is
the result of the different singularity of the Neumann function at
the corner, which is 4 times larger than that of the Green
function. At the corner there are 3 image charges --- the number
of images that one sees when standing near two perpendicular
mirror plates. In general, for a small hole located at a corner of
an opening angle $\alpha$ (see Fig. \ref{f:angle}), the MFPT is to
leading order
\begin{equation}
\label{eq:MFPT-corner} E\tau =
\frac{|\Omega|}{D\alpha}\left(\log\frac{1}{\varepsilon} + O(1)
\right).
\end{equation}
This result is a consequence of the method of images for integer
values of $\ds{\frac{\pi}{\alpha}}$. For non-integer
$\ds{\frac{\pi}{\alpha}}$ we use the complex mapping $z \mapsto
z^{\pi/\alpha}$ that flattens the corner. The upper half plane
Neumann function $\ds{\frac{1}{\pi}}\log z$ is mapped to
$\ds{\frac{1}{\alpha}}\log z$ and the analysis of Section
\ref{sec:leading} gives (\ref{eq:MFPT-corner}).

{\bf To see that the area factor $|\Omega|$ remains unchanged
under the conformal mapping $f:(x,y)\mapsto (u(x,y),v(x,y))$, we
note that this factor is a consequence of the compatibility
condition, that relates the area to the integral
$$\int_{\Omega} \Delta_{(x,y)}w\,dx\,dy = -\ds{\frac{|\Omega|}{D}},$$
where $w(x,y)=E[\tau|\,x(0)=x,y(0)=y]$ satisfies $\Delta_{(x,y)}w
= -1/D$. The Laplacian transforms according to
$$\Delta_{(x,y)} w = (u_x^2+u_y^2)\Delta_{(u,v)}w,$$
by the Cauchy-Riemann equations and the Jacobian of the
transformation is $J = u_x^2+u_y^2$. Therefore,
 $$\int_{\Omega} \Delta_{(x,y)}w\,dx\,dy =
 \int_{f(\Omega)}\Delta_{(u,v)}w\,du\,dv.$$
This means that the compatibility condition of Section
\ref{sec:leading} remains unchanged and gives the area of the
original domain.}

\section{Domains with cusps}
Here we find the leading order term of the MFPT for small holes
located near a cusp of the boundary. A cusp is a singular point of
the boundary. As $\alpha=0$ at the cusp, one expects to find a
different asymptotic expansion than (\ref{eq:MFPT-corner}). As an
example, consider the Brownian motion inside the domain bounded
between the circles $(x-1/2)^2+y^2=1/4$ and $(x-1/4)^2+y^2=1/16$
(see Figure \ref{f:cusp}). The conformal mapping $z \mapsto
\exp\{\pi i (1/z-1) \}$ maps this domain onto the upper half plane.
Therefore, the MFPT is to leading order
\begin{equation}
\label{eq:MFPT-cusp} E\tau =
\frac{|\Omega|}{D}\left(\frac{1}{\varepsilon} + O(1) \right).
\end{equation}
{\bf This result can also be obtained by mapping the cusped domain
to the unit circle. The absorbing boundary is then transformed to
an exponentially small arc of length
$\exp\{-\pi/\varepsilon\}+O(\exp\{-2\pi/\varepsilon\})$, and
equation (\ref{eq:MFPT-cusp}) is recovered.}

If the ratio between the two radii is $d<1$, then the conformal
map that maps the domain between the two circles to the upper half
plane is $\exp\left\{\ds{\frac{\pi i}{d^{-1}-1}} (1/z-1) \right\}$
(for $d=1/2$ we arrive at the previous example), so the MFPT is to
leading order
\begin{equation}
\label{eq:MFPT-cusp2} E\tau =
\frac{|\Omega|}{(d^{-1}-1)D}\left(\frac{1}{\varepsilon} + O(1)
\right).
\end{equation}
The MFPT tends algebraically fast to infinity, much faster than
the $O\left(\log\ds{\frac{1}{\varepsilon}}\right)$ behavior near
smooth or corner boundaries. The MFPT for a cusp is much larger
because it is more difficult for the Brownian motion to enter the
cusp than to enter a corner. {\bf The MFPT (\ref{eq:MFPT-cusp2})
can be written in terms of $d$ instead of the area. Substituting
$|\Omega|=\pi R^2 (1-d^2)$, we find
\begin{equation}
E\tau = \frac{\pi R^2 d(1+d)}{D}\left(\frac{1}{\varepsilon} + O(1)
\right),
\end{equation}
where $R$ is the radius of the outer circle. Note that although
the area of $\Omega$ is a monotonically decreasing function of
$d$, the MFPT is a monotonically increasing function of $d$ and
tends to a finite limit as $d\to 1$.}

Similarly, one can consider different types of cusps and find that
the leading order term for the MFPT is proportional to
$1/\varepsilon^{\lambda}$, where $\lambda$ is a parameter that
describes the order of the cusp, and can be obtained by the same
technique of conformal mapping.

\section{Diffusion on a 3-sphere}
\subsection{Small absorbing cap}
Consider a Brownian motion on the surface of a 3-sphere of radius
$R$ \cite{Oksendal}, described by the spherical coordinates
$(\theta,\phi)$
 $$x=R\sin \theta \cos \phi, \quad y=R\sin \theta \sin \phi, \quad
 z=R\cos \theta.$$
The particle is absorbed when it reaches a small spherical cap. We
center the cap at the north pole, $\theta=0$. Furthermore, the FPT
to hit the spherical cap is independent of the initial angle
$\phi$, due to rotational symmetry. Let $v(\theta)$ be the MFPT to
hit the spherical cap. Then $v$ satisfies
\begin{equation}
\Delta_M v = -1,
\end{equation}
where $\Delta_M$ is the Laplace-Beltrami operator \cite{Oksendal}
of the 3-sphere. This Laplace-Beltrami operator $\Delta_M$
replaces the regular plane Laplacian, because the diffusion occurs
on a manifold \cite[and reference therein]{Oksendal}. For a
function $v$ independent of the angle $\phi$ the Laplace-Beltrami
operator is (see Appendix \ref{ap:laplace})
\begin{equation}
\Delta_M v = R^{-2}\left( v'' + \cot \theta\,
v'\right).\label{ODE}
\end{equation}
The MFPT also satisfies the boundary conditions
\begin{equation}
v'(\pi) = 0, \quad v(\delta)=0,\label{BC}
\end{equation}
where $\delta$ is the opening angle of the spherical cap. The
solution of the boundary value problem (\ref{ODE}), (\ref{BC}) is
given by
\begin{equation}
\label{eq:v-sphere} v(\theta) = 2R^2 \log\frac{\sin
(\theta/2)}{\sin(\delta/2)}.
\end{equation}
Not surprisingly, the maximum of the MFPT is attained at the point
$\theta=\pi$ with the value
\begin{equation}
v_{\mbox{max}} = v(\pi) = -2R^2\log\sin\ds{\frac{\delta}{2}} =
2R^2 \left(\log \frac{1}{\delta} + \log 2 + O(\delta^2) \right).
\end{equation}
The MFPT, averaged with respect to a uniform initial distribution,
is
\begin{eqnarray}
E\tau &=& \frac{1}{2\cos^2\ds{\frac{\delta}{2}}}\int_{\delta}^\pi
v(\theta)\sin\theta \,d\theta \nonumber \\
&&\nonumber\\
 &=& -2R^2 \left(\frac{\log \sin(\delta/2)}{\cos^2(\delta/2)}+\frac{1}{2} \right) \nonumber \\
&&\nonumber\\
 &=& 2R^2 \left(\log \frac{1}{\delta} + \log 2 -
\frac{1}{2} + O(\delta^2\log \delta) \right).
\end{eqnarray}
Both the average MFPT and the maximum MFPT are
\begin{equation}
\label{eq:manifold-no-boundary} \tau =
\frac{|\Omega|_g}{2\pi}\left(\log\frac{1}{\delta} + O(1)\right),
\end{equation}
where $|\Omega|_g=4\pi R^2$ is the area of the 3-sphere. This
asymptotic expansion is the same as for the planar problem of an
absorbing circle in a disk. The result is two times smaller than
the result (\ref{eq:holcman}) that holds when the absorbing
boundary is a small window of a reflecting boundary. The factor
two difference is explained by the aspect angle that the particle
``sees". The two problems also differ in that the ``narrow escape"
solution is almost constant and has a boundary layer near the
window, with singular fluxes near the edges, whereas in the
problem of puncture hole inside a domain the flux is regular and
there is no boundary layer (the solution is simply obtained by
solving the ODE).

\subsection{Mapping of the Riemann sphere}
We present a different approach for calculating the MFPT for the
Brownian particle diffusing on a sphere. We may assume that the
radius of the sphere is $1/2$, and use the stereographic projection
that maps the sphere into the plane \cite{Hille2}. The point
$Q=(\xi,\eta,\zeta)$ on the sphere (often called the Riemann sphere)
$$\xi^2+\eta^2+(\zeta-1/2)^2 = (1/2)^2$$ is projected to a plane
point $P=(x,y,0)$ by the mapping
\begin{equation}
x = \frac{\xi}{1-\zeta}, \quad y=\frac{\eta}{1-\zeta}, \quad
r^2=x^2+y^2=\frac{\zeta}{1-\zeta},
\end{equation}
and conversely
\begin{equation}
\xi = \frac{x}{1+r^2}, \quad \eta=\frac{y}{1+r^2}, \quad
\zeta=\frac{r^2}{1+r^2}.
\end{equation}
The stereographic projection is conformal and therefore transforms
harmonic functions on the sphere harmonic functions in the plane,
and {\em vice versa}. However, the stereographic projection is not
an isometry. The Laplace-Beltrami operator $\Delta_M$ on the
sphere is mapped onto the operator $(1+r^2)^2 \Delta$ in the plane
($\Delta$ is the Cartesian Laplacian). The decapitated sphere is
mapped onto the interior of a circle of radius
\begin{equation}
r_\delta = \cot \frac{\delta}{2}.
\end{equation}
Therefore, the problem for the MFPT on the sphere is transformed
into the planar Poisson radial problem
\begin{equation}
\Delta V = -\frac{1}{(1+r^2)^2}, \quad\mbox{for}\quad r<r_\delta,
\end{equation}
subject to the absorbing boundary condition
\begin{equation}
V(r=r_\delta) = 0,
\end{equation}
where
 \[V(r)=v(\theta).\]
The solution of this problem is
\begin{equation}
\label{eq:sphere-all-absorbing} V(r) = \frac{1}{4} \log
\left(\frac{1+r_\delta^2}{1+r^2} \right).
\end{equation}
Transforming back to the coordinates on the sphere, we get
\begin{equation}
\label{eq:v-Riemann}  v(\theta) = \frac12 \log\frac{\sin
(\theta/2)}{\sin(\delta/2)}.
\end{equation}
As the actual radius of the sphere is $R$ rather than $1/2$,
multiplying eq.(\ref{eq:v-Riemann}) by $(2R)^2$, we find that
(\ref{eq:v-Riemann}) is exactly (\ref{eq:v-sphere}).

\subsection{Small cap with an absorbing arc}
\begin{figure}
\includegraphics{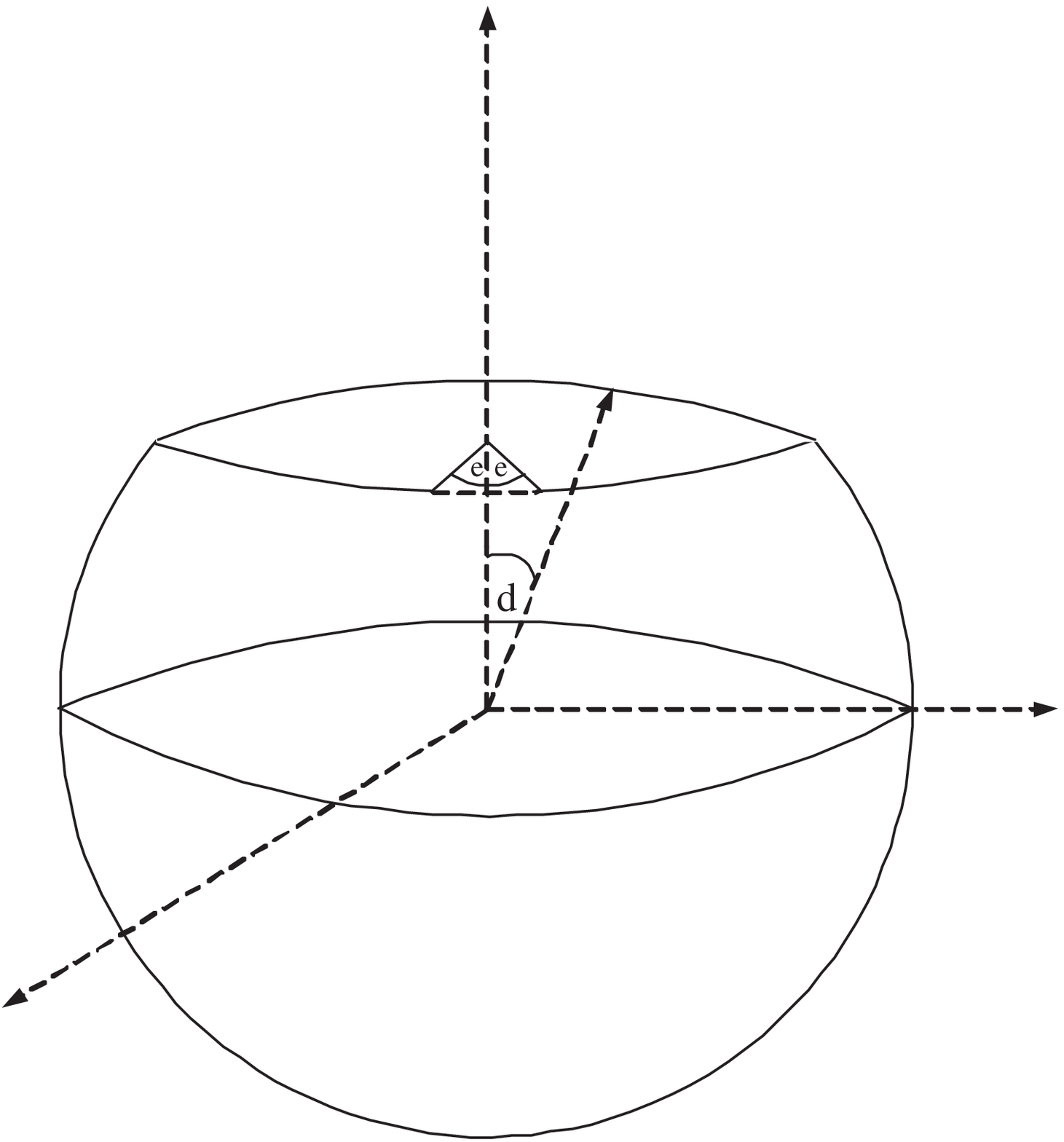}
\caption{A sphere of radius $R$ without a spherical cap at the
north pole of central angle $\delta$. The particle can exit
through an arc seen at angle $2\varepsilon$.}\label{f:sphere}
\end{figure}

Consider again a Brownian particle diffusing on a decapitated
3-sphere of radius $1/2$. The boundary of the spherical cap is
reflecting but for a small window that is absorbing (see Fig.
\ref{f:sphere}). We calculate the mean time to absorption. Using
the stereographic projection of the preceding subsection, we
obtain the mixed boundary value problem
\begin{eqnarray}
\Delta v &=& -\frac{1}{(1+r^2)^2}, \quad\mbox{for}\quad
r<r_\delta, \quad 0\leq \phi < 2\pi,\nonumber
\\
&&\nonumber\\
 v(r,\phi)\bigg|_{r=r_\delta} &=& 0, \quad\mbox{for}\quad
|\phi-\pi|<\varepsilon, \nonumber\\
&&\nonumber\\
\frac{\partial v(r,\phi)}{\partial r}\bigg|_{r=r_\delta} &=& 0,
\quad\mbox{for}\quad |\phi-\pi|>\varepsilon.
\end{eqnarray}
The function
 $$w(r) = \frac{1}{4} \log
 \left(\ds{\frac{1+r_\delta^2}{1+r^2}} \right)$$
is the solution of the all absorbing boundary problem
eq.(\ref{eq:sphere-all-absorbing}), so the function $u=v-w$
satisfies the mixed boundary value problem
\begin{eqnarray}
\Delta u &=& 0, \quad r<r_\delta, \quad\mbox{for}\quad 0\leq \phi
< 2\pi,\nonumber\\
&&\nonumber\\
u(r,\phi)\bigg|_{r=r_\delta} &=& 0, \quad\mbox{for}\quad
|\phi-\pi|<\varepsilon, \nonumber\\
&&\nonumber\\
\frac{\partial u(r,\phi)}{\partial r}\bigg|_{r=r_\delta} &=&
\frac{r_\delta}{2(1+r_\delta^2)}, \quad\mbox{for}\quad
|\phi-\pi|>\varepsilon.
\end{eqnarray}
Scaling $\tilde{r}=r/r_\delta$, we find this mixed boundary value
problem to be that of a planar disk \cite{NarrowEscape2}, with the
only difference that the constant $1/2$ is now replaced by
$\ds{\frac{r_\delta^2}{2(1+r_\delta^2)}}$. Therefore, the solution
is given by
\begin{equation}
a_0 =
-\frac{2r_\delta^2}{1+r_\delta^2}\left[\log\frac{\varepsilon}{2} +
O(\varepsilon)\right].
\end{equation}
Transforming back to the spherical coordinate system, the MFPT is
 \beq
v(\theta,\phi) &=&
\frac{1}{2}\log\frac{\sin\theta/2}{\sin\delta/2}\nonumber\\
&&\label{eq:MFPT-sphere-north}\\
&&-\cos^2\frac{\delta}{2}\left[\log\frac{\varepsilon}{2} +
O(\varepsilon)\right] + \sum_{n=1}^\infty a_n \left[\frac{\cot
\left(\theta/2\right)}{\cot \left(\delta/2\right)}\right]^n\cos
n\phi.\nonumber
 \eeq
The MFPT, averaged over uniformly distributed initial conditions
on the decapitated sphere, is
\begin{equation}
E\tau = -\frac{1}{2} \left(\frac{\log \sin(\delta/2)}{\cos
^2(\delta/2)}+\frac{1}{2} \right) +
\cos^2\frac{\delta}{2}\left[\log\frac{2}{\varepsilon} +
O(\varepsilon)\right].\label{Etau}
\end{equation}
Scaling the radius $R$ of the sphere into (\ref{Etau}), we find
that for small $\varepsilon$ and $\delta$ the averaged MFPT  is
\begin{equation}
E\tau = 2R^2
\left[\log\frac{1}{\delta}+2\log\frac{1}{\varepsilon}+3\log 2 -
\frac{1}{2} +
O(\varepsilon,\delta^2\log\delta,\delta^2\log\varepsilon) \right].
\end{equation}
There are two different contributions to the MFPT. The ratio
$\varepsilon$ between the absorbing arc and the entire boundary
brings in a logarithmic contribution to the MFPT, which is to
leading order $$\frac{|\Omega|_g}{\pi}\log\frac{1}{\varepsilon}.$$
However, the central angle $\delta$ gives an additional
logarithmic contribution, of the form
$$\frac{|\Omega|_g}{2\pi}\log\frac{1}{\delta}.$$ The factor 2
difference in the asymptotic expansions is the same as encountered
in the planar annulus problem.

The MFPT for a particle initiated at the south pole $\theta=\pi$
is
\begin{eqnarray}
v(\pi) &=& -2R^2 \log\sin\ds{\frac{\delta}{2}} -
4R^2\cos^2\frac{\delta}{2}\left[\log\frac{\varepsilon}{2} +
O(\varepsilon)\right] \nonumber \\
&&\nonumber\\
 &=& 2R^2 \left[\log\frac{1}{\delta} + 2\log
\frac{1}{\varepsilon} + 3\log 2 +
O(\varepsilon,\delta^2\log\delta,\delta^2\log\varepsilon)\right].
\end{eqnarray}
We also find the location $(\theta,\phi)$ for which the MFPT is
maximal. The stationarity condition $\ds{\frac{\partial
v}{\partial \phi}}=0$ implies that $\phi=0$, as expected (the
opposite $\phi$-direction to the center of the window). The
infinite sum in equation (\ref{eq:MFPT-sphere-north}) is $O(1)$.
Therefore, for $\delta \ll 1$, the MFPT is maximal near the south
pole $\theta=\pi$. However, for $\delta=O(1)$, the location of the
maximal MFPT is more complex.

Finally, we remark that the stereographic projection also leads to
the determination of the MFPT for diffusion on a 3-sphere with a
small hole as discussed above, and an all reflecting spherical cap
at the south pole. In this case, the image for the stereographic
projection is the annulus, a problem solved in Section
\ref{sec:annulus}.

\appendix
\section{Laplace Beltrami operator on 3-sphere}
\label{ap:laplace} The Laplace Beltrami operator on a manifold is
given by
\begin{equation}
\Delta_M f = \frac{1}{\sqrt{\det G}}\sum_{i,j} \frac{\partial
}{\partial \xi_i} \left(g^{ij} \sqrt{\det G}\frac{\partial
f}{\partial \xi_j} \right),
\end{equation}
where
\begin{equation}
\mb{t}_i = \frac{\partial \mb{r}}{\partial \xi_i}, \quad g_{ij} =
\langle \mb{t}_i, \mb{t}_j \rangle, \quad G = (g_{ij}), \quad
g^{ij}=g_{ij}^{-1}.
\end{equation}
In spherical coordinates we have
\begin{equation}
g_{\theta \theta} = R^2, \quad g_{\phi \phi} = R^2 \sin ^2 \theta,
\quad g_{\theta \phi}=g_{\phi \theta} = 0.
\end{equation}
Therefore, for a function $w=w(\theta,\phi)$
\begin{equation}
\Delta_Mf = R^{-2} \left(\frac{\partial^2 f}{\partial \theta^2} +
\cot \theta \, \frac{\partial f}{\partial \theta} +
\frac{1}{\sin^2\theta}\frac{\partial ^2 f}{\partial
\phi^2}\right).
\end{equation}

\noindent {\bf Acknowledgment:} This research was partially
supported by research grants from the Israel Science Foundation,
US-Israel Binational Science Foundation, and the NIH Grant No.
UPSHS 5 RO1 GM 067241.

\end{document}